\documentclass[12pt]{article}%
\usepackage{amsmath}
\usepackage{amsfonts}
\usepackage{amssymb}
\usepackage{graphicx}%
\newtheorem{theorem}{Theorem}

\newenvironment{proof}[1][Proof]{\noindent\textbf{#1.} }{\ \rule{0.5em}{0.5em}}

\begin{document}

\title{ 
Calculation of the Unitary part of the Bures Measure for N-level Quantum Systems }

\author{Renan Cabrera, Herschel Rabitz\\
\small Department of Chemistry, Princeton University, Princeton, New Jersey 08544, USA \\
\small rcabrera@princeton.edu
}

\maketitle 
\begin{abstract}
We use the canonical coset parametrization and provide a formula with the unitary 
part of the Bures measure for non-degenerate systems in terms of the product 
of even Euclidean balls. This formula is shown to be consistent
with the sampling of random states through the generation of random 
unitary matrices.
\end{abstract}

\maketitle 

\emph{\small This is an author-created, un-copyedited version of an article accepted for publication in J. Phys. A: Math. Theor. IOP Publishing Ltd is not responsible for any errors or omissions in this version of the manuscript or any version derived from it. The definitive publisher authenticated version is available online 
at  10.1088/1751-8113/42/44/445302 }

\section{Introduction}
The Bures measure is the volume element of the corresponding Bures metric that can be
obtained from the infinitesimal form of the quantum fidelity between mixed 
quantum states \cite{uhlmann1992mba,nielsenchuang,sommers2003bvs}, or from 
the statistical distance between mixed quantum states \cite{PhysRevLett.72.3439}. 
The  Bures measure has been proposed as a prior distribution for implementation of quantum 
Bayes estimation \cite{slater1996qfb,Slater19981}. Other proposals and other measures
also exist  including monotonic Riemannian measures \cite{petz:2662,Slater19981}. 
The term monotonic is applied to stochastic maps, which are not allowed to increase 
the distance.
Andai \cite{andai2006volume} calculated the volume of the whole state space according to 
the Lebesgue measure including a few monotonic Riemannian measures \cite{petz:2662}.

The quadratic form of the Bures metric can be written as 
\begin{equation}
  dB(\rho,\rho+d\rho)^2 = \frac{1}{2}Tr[ G d\rho    ],
\end{equation}
with $G$ implicitly defined from $d\rho = G\rho + \rho G $. A more practical formula was found 
by H\"ubner \cite{Hubner1992239} in terms of the eigenvalues (populations) of 
the state as 
\begin{equation}
  dB(\rho,\rho+d\rho)^2 = \frac{1}{2}\sum_{j=1}^N \sum_{k=1}^N 
 \frac{ | \langle \lambda_j | d\rho | \lambda_k \rangle |^2 }{\lambda_j +\lambda_k},
\label{hubner-eq}
\end{equation}
where $ \rho | \lambda_k \rangle = \lambda_k  |\lambda_k \rangle $. A general state can be parametrized by applying a 
unitary operator on the diagonal state  $\rho^{(D)} =  diag( \lambda_1, \lambda_2,...\lambda_N  ) $ as
\begin{equation}
\rho  = \Omega \rho^{(D)} \Omega^\dagger,
\end{equation}
with the unitary operator  \cite{byrd2007gns,1751-8121-40-37-010} as the
generalized flag manifold  
\begin{equation}
\Omega \in  \frac{U(N)}{U(m_1)\otimes U(m_2) \otimes ...U(m_q)},\,\,\, m_1+m_2+...m_q=N 
\end{equation}
where $m_j$ is the degeneracy of the unique eigenvalue $\lambda_j$. The flag 
manifold can be  decomposed as a product of cosets in order to develop a 
suitable parametrization. For example, the 
non-degenerate case can be decomposed as   
\begin{equation}
\Omega \in  \frac{U(N)}{U(1)^{\otimes N}} =  
 \frac{U(N)}{U(N-1)\otimes U(1)} \,\, \frac{U(N-1)}{U(N-2)\otimes U(1)}...\frac{U(2)}{U(1)\otimes U(1)}.
\label{main-coset} 
\end{equation}
The eigenstates can be written as $| \lambda_k \rangle = \Omega | k \rangle $ and the infinitesimal 
variation of $\rho$ can be expanded  as
\begin{equation}
d\rho =  \Omega d\rho^{(D)} \Omega^{\dagger} + d\Omega \rho^{(D)} \Omega^\dagger + 
  \Omega \rho^{(D)} d \Omega^\dagger.
\end{equation}
Introducing this expression in H\"ubner's formula (\ref{hubner-eq}) we find
\begin{eqnarray}
   dB(\rho,\rho+d\rho)^2 &=&
   \frac{1}{4}\sum_{j=1}^N 
    \frac{ | \langle j | d \rho^{(D)} | j \rangle|^2 }{\lambda_j}
 + \sum_{j=1}^N \sum_{k=j+1}^N  
    \frac{ |\langle j | [\Omega^\dagger d\Omega, \rho^{(D)}] | k \rangle|^2 }{ \lambda_j +\lambda_k  } \\
  &=&   \frac{1}{4}\sum_{j=1}^N 
    \frac{ (d\lambda_j)^2 }{ \lambda_j } + 
   \sum_{j=1}^N \sum_{k=j+1}^N \Lambda_{jk} |\Omega^\dagger d\Omega|^2_{jk},
\end{eqnarray}
with $\Lambda_{jk} = \frac{(\lambda_j - \lambda_k)^2}{\lambda_j+\lambda_k}$, such that the volume element can be extracted to obtain
Hall's formula \cite{Hall1998123} up to a scale factor. The volume element
with the scale used in \cite{bengtsson2006gqs} is
\begin{equation}
 dV_B =\delta( \lambda_1 + \lambda_2 + ...\lambda_N -1  ) 
\frac{d\lambda_1 d\lambda_2...d\lambda_N }{ 2^{N-2} \sqrt{\lambda_1 \lambda_2 ...\lambda_N} }
  \prod_{j<k}^N \Lambda_{jk} dx_{jk} dy_{jk},
\end{equation}
where $dx_{jk} = Re (\Omega^\dagger d\Omega)_{jk}  $ and $dy_{jk} = Im (\Omega^\dagger d\Omega)_{jk}  $. The remarkable feature of this 
expression is that it separates the effect of the eigenvalues 
(populations) and the effect of the unitary operator, 
\begin{equation}
 dV_B = dV_B^{(\lambda)} dV_B^{(\Omega)}.
\end{equation}
This paper is organized as follows. Section 2 
reviews some results of the Euler parametrization and
introduces some concepts and formulas to be used later in the paper. The main 
contribution of this paper is developed in section 3. Section 4 compares the 
results with the generation of random unitary matrices. Section 5 presents 
concluding remarks.

\section{Euler Parametrization}
The generalized Euler parametrization was developed by Tilma and collaborators
 \cite{tilma2002pbs,tilma2002gea} and was used in \cite{Tilma2004263} to 
calculate the volume and measure of the unitary part of the Bures measure. 
In this section we review the three-level case in order to introduce some concepts 
and formulas that will be used in the next section. The unitary operator for a 
three-level system can be parametrized as 
\cite{tilma2002gea,Slater2001207,slater1996qfb}
\begin{eqnarray}
   \Omega &=& ( e^{\phi^6 \lambda_3} e^{\phi^5 \lambda_2}  e^{\phi^4 \lambda_3}  e^{\phi^3 \lambda_5}) 
     ( e^{\phi^2 \lambda_3} e^{\phi^1 \lambda_2} ),
\end{eqnarray}
with the factors of $\Omega$  as parametrisation of $ \frac{U(3)}{ U(2)\otimes U(1)} $ and $ \frac{U(2)}{ U(1)\otimes U(1)} $ in terms of
the Euler angles $\phi^k$ and the Gell-Mann matrices $\lambda_k$. 
The measure of the unitary portion can be calculated as the product of 
the measure of the corresponding cosets. For non-degenerate 3-level 
systems, the coset decomposition is
\begin{equation}
 dV_{B}^{(\Omega)} =
  dV \left( \frac{U(3)}{ U(2)\otimes U(1)} \right) dV \left( \frac{U(2)}{ U(1)\otimes U(1)} \right) 
\end{equation}
The volume of the coset  $ \frac{U(3)}{ U(2)\otimes U(1)} $ can be obtained by 
calculating $\Omega^\dagger d\Omega$ and selecting the matrix components where 
the corresponding Lie algebra lies. 
Thus, we may extract the coordinate transformation from the 
following terms 
\begin{equation}
\Omega^\dagger d\Omega \equiv \begin{pmatrix} \cdot & \cdot &  {dx^3}^\prime + i {dx^4}^\prime \\
                  \cdot & \cdot &  {dx^5}^\prime + i {dx^6}^\prime \\
                  \cdot & \cdot  & \cdot 
  \end{pmatrix}
\end{equation}
to obtain
\begin{equation}
 dV \left( \frac{U(3)}{ U(2)\otimes U(1)} \right) = 
 \cos \phi^3 \sin^3 \phi^3 \sin 2\phi^5 \, d\phi^3  d\phi^4  d\phi^5  d\phi^6. 
\end{equation}
A similar procedure can be carried out for the second coset. 

The Haar measure of the coset $\frac{U(n+1)}{ U(n)\otimes U(1)}$ is topologically equivalent to 
an even sphere $S^{2n}$ according to Gilmore \cite{gilmore1973lgl,gilmore2008lgp} 
with the corresponding volume
\begin{equation}
 \text{Vol}_{Haar}\left( \frac{U(n+1)}{ U(n)\otimes U(1)}  \right) = \text{Vol}( S^{2n} )
\end{equation}
However, the measure of the unitary section of the Bures measure is not 
the Haar measure. Some references refer to it as the \emph{truncated} 
Haar Measure \cite{tilma2002gea,Byrd2001152}. 
Direct integration  of the coset measure does not result 
in the volume of even spheres $r^2=1$, but instead in the volume of even balls $r^2\le 1$,
with $r$ as the radial coordinate, such that
\begin{equation}
  \text{Vol}\left( \frac{U(n+1)}{ U(n)\otimes U(1)}  \right) = \text{Vol}( B^{2n} ),
\label{cosetCPn}
\end{equation}
with  
\begin{equation}
  \text{Vol}( B^{n} ) = \frac{2 \pi^{n/2}}{ n \Gamma(n/2) },
\end{equation}
which is in perfect agreement with the formulas found in \cite{Boya2003401,Tilma2004263}. 

Consequently, the volume of the unitary section for non-degenerate systems 
is equal to the product of the volume of even balls
\begin{equation}
 \text{Vol} \left( \frac{U(N)}{U(1)^{\otimes N}} \right) = 
 \text{Vol}(B^{2N-2})  \text{Vol}(B^{2N-4}) ...  \text{Vol}(B^{2}) 
  =  \frac{\pi^{N(N-1)/2}}{ \prod_{1}^{N} \Gamma(n) }
\label{volflag}.
\end{equation}
This result is also consistent with the volume  presented by 
Sommers and {\.Z}yczkowski \cite{sommers2003bvs} as
\begin{equation} 
  \text{Vol}^\prime \left( \frac{U(N)}{U(1)^{\otimes N}} \right) 
 = \frac{(2\pi)^{N(N-1)/2}}{ \prod_{1}^{N} \Gamma(n) }.
\end{equation}
The discrepancy factor can be explained by a simple numerical scale 
factor of $1/2$ on the Bures metric, because $N(N-1)$ is equal to the 
dimension of the Lie algebra occupied by the coset space. 
Equivalently, $N(N-1)$ is the number of degrees 
of freedom required to parametrize $ \frac{U(N)}{U(1)^{\otimes N}}$. So far, we have shown that 
the volume of the coset (\ref{cosetCPn}) can be written as the volume of 
an even ball, but this does not imply that the measure of the coset
is Euclidean defined on an even ball. This assertion is proved 
in the next section and further numerical tests are carried out in section 4.

\section{Canonical Coset Parametrization}
An important parametrization arises from the canonical coset, 
as presented by Gilmore \cite{gilmore1973lgl} on page 351. The Bures metric was obtained
for 3-level systems in \cite{akhtarshenas:012102} and a more general prescription in 
\cite{1751-8121-40-37-010} for N-level systems, but the measure was not calculated 
in this formulation. The power of the canonical coset parametrization lies in the 
many possibilities to analytically express the exponential of the following typical 
block matrix 
\begin{equation}
 \exp{ \begin{pmatrix} \mathbf{0} & B \\ -B^\dagger &0 \end{pmatrix}  } =
 \begin{pmatrix} \cos \sqrt{BB^\dagger} &  \frac{\sin \sqrt{B^\dagger}B}{\sqrt{B^\dagger}B  } B
 \\ -  \frac{\sin \sqrt{B^\dagger}B}{ \sqrt{B^\dagger}B  }  B^\dagger & 
   \cos \sqrt{B^\dagger B}  \end{pmatrix} 
\end{equation}
where the case of interest is such that $B = column ( z^1, z^2, ..., z^{N-1}  )  $
is a column vector of complex numbers. This exponential can also be expressed
in terms of spherical coordinates $x^j$ as
\begin{equation}
   \exp{ \begin{pmatrix} \mathbf{0} & B \\ -B^\dagger &0 \end{pmatrix}  } 
= \begin{pmatrix} [\mathbf{1} - XX^\dagger]^{1/2} & X \\
   -X^\dagger &  [1 - X^\dagger X]^{1/2}  \end{pmatrix}.
\label{exp-coset}
\end{equation}
such that 
\begin{equation}
 X =  \frac{\sin \sqrt{B^\dagger}B}{\sqrt{B^\dagger}B  } B =
 \begin{pmatrix}
  x^1 + i x^2 \\ x^3 +i x^4\\ \vdots \\ x^{2N-3} + i x^{2N-2}
 \end{pmatrix}
\end{equation}
This coordinate system is called spherical because the column vector
is made of variables that range inside an even ball $B^{2k}$, where the radial
coordinate is $r^2 = X^\dagger X $. The exponential in (\ref{exp-coset}) is important because it 
provides a parametrization of the coset $ \frac{U(N)}{U(N-1)\otimes U(1)}$ as a $ N \times N$ matrix. 
The coset required to parametrize the unitary section of the Bures 
measure can be constructed in terms of products of layered cosets 
(\ref{main-coset}) that have form 
{\footnotesize
\begin{eqnarray}
    \frac{U(2)}{U(1)\otimes U(1)} &=& 
    \begin{pmatrix} \sqrt{1-(x^1)^2-(x^2)^2} &  x^1+ i x^2 &0 &0... & 0  \\
                    -(x^1-ix^2) & \sqrt{1-(x^1)^2-(x^2)^2}          &0 &0... & 0  \\
                    0           & 0          &1 &0... & 0  \\
                    \vdots     & 0           &0 &1... & 0  \\
                    0          &0            &0 &0... & 1  
     \end{pmatrix} \nonumber \\
   \frac{U(3)}{U(2)\otimes U(1)} &=& 
    \begin{pmatrix} W_{11}^{(2)}       &  W_{12}^{(2)} &x^3+ix^4& 0... & 0  \\
                    W_{21}^{(2)}       & W_{22}^{(2)}          &x^5+ix^6& 0... & 0  \\
                    -(x^3-ix^4) & -(x^5-ix^6)           &\sqrt{1-(x^3)^2-..}& 0... & 0  \\
                    \vdots      & 0          &0& 1... & 0  \\
                    0           & 0          &0& 0... & 1  
     \end{pmatrix}\nonumber\\
\vdots,  \nonumber
\end{eqnarray}}with $W_{jk}^{(n)} = (\sqrt{ \mathbf{1} - X X^\dagger  })_{jk}$. With this background, we state the following theorem
\begin{theorem}
The measure of the following coset corresponds to an Euclidean measure 
defined inside of an even ball, such that

\begin{equation}
 dV \left( \frac{U(n+1)}{ U(n)\otimes U(1)}  \right) = dV_E(B^{2n}).
\label{unitarybures}
\end{equation}

\end{theorem}

This theorem applied to 3-level systems results in
\begin{equation}
  dV \left( \frac{U(3)}{U(2)\otimes U(1)} \right) = dx^3 dx^4 dx^5 dx^6, 
\end{equation}
such that $ (x^3)^2 + (x^4)^2 +   (x^5)^2 +  (x^6)^2 \le 1 $, in terms of the variables 
of the corresponding canonical coset parametrization. 

\begin{proof}
The strategy is based in the generalization of the proof initially provided 
for the simpler case  $dV \left( \frac{U(3)}{U(2)\otimes U(1)} \right)$. The complex column $X$ of interest is 
\begin{equation}
 X =  \begin{pmatrix}
  x^3 + i x^4 \\ x^5 +i x^6
 \end{pmatrix},
\end{equation}
such that the unitary operator becomes
\begin{equation}
\Omega = \begin{pmatrix} [\mathbf{1} - XX^\dagger]^{1/2} & X \\
   -X^\dagger &  [1 - X^\dagger X]^{1/2}  \end{pmatrix} = 
 \begin{pmatrix} \mathbf{1} + \frac{\sqrt{1-r^2}-1}{r^2}X X^\dagger & X \\
   -X^\dagger & \sqrt{1-r^2}
 \end{pmatrix}, 
\end{equation}
with $r^2 = X^\dagger X = (x^3)^2 +(x^4)^2 + (x^5)^2 + (x^6)^2 $. The measure is invariant
under an orthonormal transformation $\mathcal{O}$ applied to the coordinates $( x^3,x^4,x^5,x^6)$.
This means that it is sufficient to consider the evaluation of  $ \Omega^\dagger d\Omega$ at 
 $( x^3,x^4,x^5,x^6) = (r,0,0,0)$, which produces the following expression
\begin{equation}
 \Omega^\dagger d\Omega = 
\begin{pmatrix}
-i r dx^4 & -\frac{(\sqrt{1-r^2}-1)(dx^5-i dx^6)}{r} & \frac{dx^3-i(r^2-1)dx^4}{\sqrt{1-r^2}}\\
\frac{(\sqrt{1-r^2}-1)(dx^5+i dx^6)}{r} & 0 & dx^5 +i dx^6  \\
-\frac{dx^3+i(r^2-1)dx^4}{\sqrt{1-r^2}} & -dx^5 +i dx^6 & i r dx^4
\end{pmatrix}.
\end{equation}
The coordinate transformation can be extracted from $(\Omega^\dagger d\Omega)_{13}$ and 
$(\Omega^\dagger d\Omega)_{23}$, as
\begin{equation}
\begin{pmatrix} {dx^3}^\prime \\ {dx^4}^\prime \\  {dx^5}^\prime \\  {dx^6}^\prime \end{pmatrix} 
  = 
\begin{pmatrix}
  \frac{1}{\sqrt{1-r^2}} &0&0&0\\
  0& \sqrt{1-r^2} &0 &0 \\
  0 & 0 & 1 &0 \\
  0&0&0&1
\end{pmatrix} 
 \begin{pmatrix} dx^3 \\ dx^4 \\  dx^5 \\  dx^6 \end{pmatrix},  
\label{transf3} 
\end{equation}
which leads to the measure
\begin{equation}
  dV \left( \frac{U(3)}{U(2)\otimes U(1)} \right) = dx^3 dx^4 dx^5 dx^6, 
\end{equation}
with $r\le 1$. The transformation matrix (\ref{transf3}) changes  with the application
of an orthonormal operator $\mathcal{O}$ on the coordinates $( x^3,x^4,x^5,x^6)$, but the determinant 
remains invariant as stated before. The only differentials without trivial 
transformation are $dx^3$ and $dx^4$ which were evaluated at $x^3=r$ and $x^4=0$.
The rest of the differentials transform according to the identity. This means that 
by extending the coordinates to higher  dimensions the extra differentials
will transform according to the identity as well, thus, maintaining the determinant 
equal to $1$ and proving the theorem.
\end{proof}

This theorem leads us to formulate the following formula of the Bures measure 
for a state with non-degenerate spectrum  
\begin{equation}
  dV_B^{(\Omega)}(N) =   dV(B^{2N-2})dV(B^{2N-4})...dV(B^2).
\end{equation}

Some degenerate states including pure states and those without full-rank 
can be treated by reducing the degrees of freedom and the number of balls 
involved in the parametrization. For example, Table \ref{Table:degen-bures} 
shows the characteristic diagonal states along with their corresponding  
measures in low dimensions.
\begin{table}[ht]
\centering
\begin{tabular}{|l|c|l|}
\hline
 Diagonal state  & $\Omega$ & Measure \\
\hline
 $diag(0,0,1)$ & $\frac{U(3)}{U(2)\otimes U(1)}$ & $dV(B^{4})$ \\
  $diag(0,0,0,1)$ & $\frac{U(4)}{U(3)\otimes U(1)}$ & $dV(B^{6})$ \\

  $diag(0,0,\lambda_2,\lambda_1)$ &
   $\frac{U(4)}{U(3)\otimes U(1)} \frac{U(3)}{U(2)\otimes U(1)}  $ & $dV(B^{6}) dV(B^{4})$ \\

  $diag(0,0,0,0,1)$ &
   $\frac{U(5)}{U(4)\otimes U(1)}  $ & $dV(B^{8})$ \\

  $diag(0,0,0,\lambda_2,\lambda_1)$ &
   $\frac{U(5)}{U(4)\otimes U(1)} \frac{U(4)}{U(3)\otimes U(1)}  $ & $dV(B^{8})dV(B^{6})  $ \\

  $diag(0,0,\lambda_3,\lambda_2,\lambda_1)$ &
   $\frac{U(5)}{U(4)\otimes U(1)} \frac{U(4)}{U(3)\otimes U(1)} 
   \frac{U(3)}{U(2)\otimes U(1)} $ & $dV(B^{8}) dV(B^{6}) dV(B^{4})  $ \\
\hline
\end{tabular}
\caption{ Unitary part of the Bures measure for degenerate states with reduced rank, 
 where $\lambda_j \neq \lambda_k $.  }
\label{Table:degen-bures}
\end{table}

\section{Random Sampling}
\begin{figure} 
\centering
\includegraphics[scale=0.7]{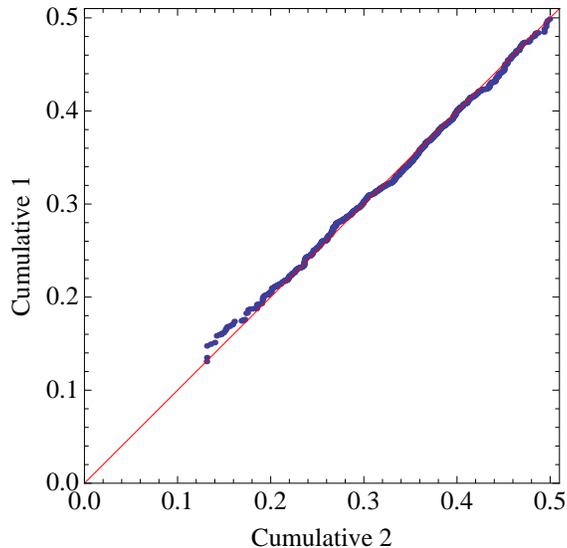}
\caption{ Plot of the equivalent cumulatives of the component $(\rho)_{33}$ for 
1000 random states with spectrum  $diag \left( \frac{3}{8} , \frac{1}{8} , \frac{1}{2}  \right)$. }
\label{fig:cumulative1}
\end{figure}

The results from the previous sections can be used to compare
the sampling distribution of the Euclidean spheres of the Bures measure 
against the sampling distribution obtained from a generation of random unitary matrices. 
The most efficient and transparent method to generate random unitary matrices 
is described by Mezzadri \cite{mezzadri2007grm}, which is based on the 
QR decomposition of complex random matrices. A random state $\rho$ can be generated 
by two independent methods
\begin{itemize}
 \item 1: Through the generation of random unitary matrices.
 \item 2: Through the generation of random points on the even Euclidean balls $B^{2k}$
 and subsequent  use of the canonical coset parametrization to obtain the state.
\end{itemize}  
The alternative distributions seem to be equivalent as can be verified by plotting 
their cumulatives against each other and observing a linear one-to-one correspondence 
up to some fluctuations. A specific test can be designed for states having the 
spectrum of the following non-degenerated diagonal state
\begin{equation}
 \rho^{(D)} = diag \left( \frac{3}{8} , \frac{1}{8} , \frac{1}{2}  \right).
\end{equation}
A plot of the two cumulatives against each other for the $(\rho)_{33}$ component is 
shown in Figure \ref{fig:cumulative1}. This test was comprehensively 
carried out and verified for systems up to 5-levels.

\section{Conclusions}

We calculated the unitary part of the Bures measure for non-degenerate
systems in terms of the canonical coset parametrization and found 
an expression as the product of Euclidean even balls. 
This result was shown to be in agreement with the numerical random sampling of 
unitary matrices and with the formulas of the volume found 
in the literature. These results are also relevant to monotone metrics other 
than the Bures metric including the Hilbert-Schmidt measure.

\section*{Acknowledgment}
The authors acknowledge the support from the N.S.F. and A.R.O.

\section*{References}

\bibliographystyle{unsrt}
\bibliography{BuresMeasure}

\end{document}